\documentclass[12pt]{article}
\usepackage{epsfig}
\usepackage{latexsym}
\usepackage{amstext}
\usepackage{amssymb}
\usepackage{graphics}

\newcommand{\mb}[1]{ \mbox{\boldmath$#1$}}
\newcommand{\ds}{\displaystyle}
\newcommand{\beq}{\begin{eqnarray}}
\newcommand{\eeq}{\end{eqnarray}}
\newcommand{\beqq}{\begin{eqnarray*}}
\newcommand{\eeqq}{\end{eqnarray*}}
\newcommand{\p}{\partial}

\newcommand{\eps}{\varepsilon}

\font\bb=msbm10 at 12pt

\def\eE{\hbox{\bb E}}

\begin{document}
\pagestyle{plain}
\begin{center}
{\large \textbf{{Stochastic resonance with applied and induced
fields: the case of voltage-gated ion channels}}
\\[5mm]
M. Shaked \footnote{Department of Systems, School of Electrical
Engineering, The Iby and Aladar Fleischman Faculty of Engineering,
Tel-Aviv University, Ramat-Aviv Tel-Aviv 69978, Israel.}, Z. Schuss
\footnote{Department of Mathematics, Tel-Aviv University, Tel-Aviv
69978, Israel.}}
\end{center}
\date{}

\begin{abstract}
\noindent We consider a charged Brownian particle in an asymmetric
bistable electrostatic potential biased by an externally applied or
induced time periodic electric field. While the amplitude of the
applied field is independent of frequency, that of the one induced
by a magnetic field is. Borrowing from protein channel terminology,
we define the open probability as the relative time the Brownian
particle spends on a prescribed side of the potential barrier. We
show that while there is no peak in the open probability as the
frequency of the {\em applied} field and the bias (depolarization)
of the potential are varied, there is a narrow range of low
frequencies of the {\em induced} field and a narrow range of the low
bias of the potential where the open probability peaks. This
manifestation of stochastic resonance is consistent with
experimental results on the voltage gated $I_{\mbox{\scriptsize
Ks}}$ and KCNQ1 potassium channels of biological membranes and on
cardiac myocytes.
\end{abstract}

\section{Introduction}

\begin{figure}
\centering
{\includegraphics[width=8cm]{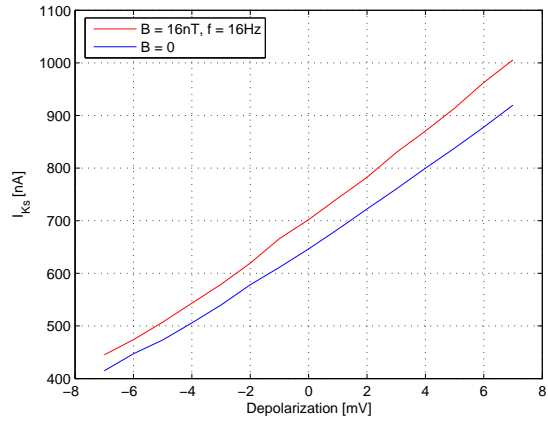}}
\caption{\small $I_{\mbox{\scriptsize Ks}}$ current in {\em Xenopus}
oocytes with applied magnetic field of 16 Hz and 16 nT (red) and
without (blue)} \label{f:redblue}
\end{figure}

\begin{figure}
\centering
{\includegraphics[width=8cm]{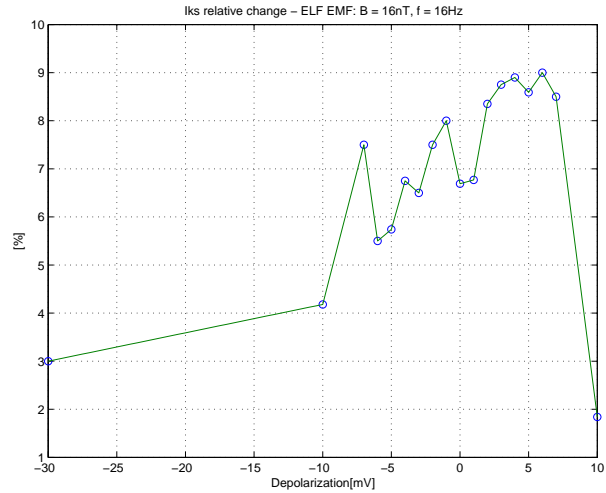}}
\caption{\small The quotient of $I_{\mbox{\scriptsize Ks}}$
expressed in {\em Xenopus} oocytes with applied magnetic field of 16
Hz and 16 nT and without (red to blue in Figure \ref{f:redblue})}
\label{f:SIKUM_Iks_change_B_16nT_f16}
\end{figure}

\begin{figure}
\centering
{\includegraphics[width=8cm]{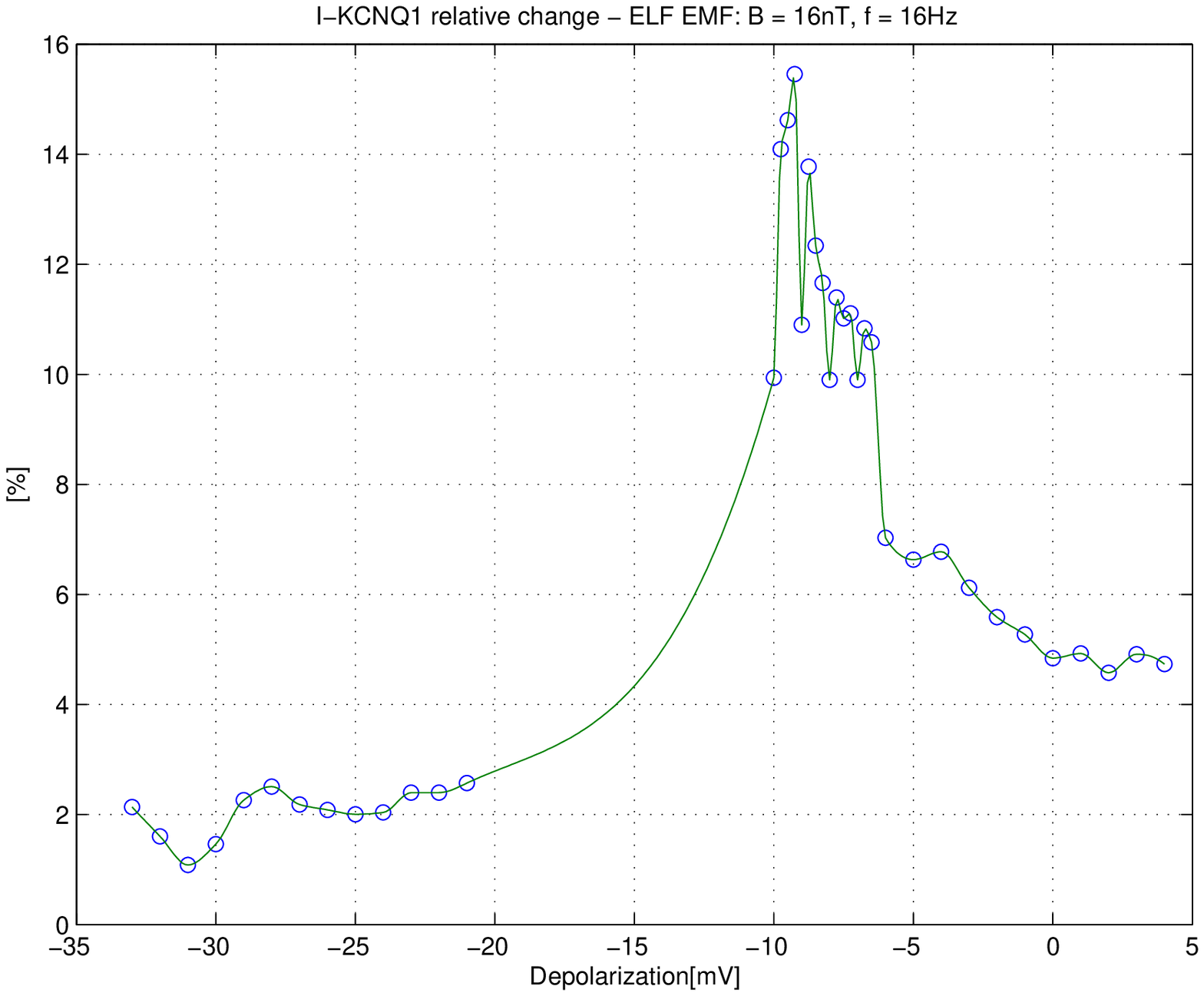}}
\caption{\small The quotient of KCNQ1 expressed in {\em Xenopus}
oocytes with applied magnetic field of 16 Hz and 16 nT and without}
\label{f:SIKUM_KCNQ1_change_B_16nT_f16}
\end{figure}

Our recent experimental findings  show unusual non-thermal
biological effects of a periodic electromagnetic field (EMF) of
frequency 16 Hz and amplitude 16 nT (nano Tesla) on the potassium
current in human $I_{\mbox{\scriptsize Ks}}$ and KCNQ1 channels
\cite{Attali}. More specifically, we expressed the
$I_{\mbox{\scriptsize Ks}}$ channel in {\em Xenopus} oocytes and
varied the membrane depolarization between -100 mV and +100 mV and
measured the membrane potassium current. The current with applied
EMF peaked above that without applied EMF at membrane
depolarizations between 0 mV and 8 mV to a maximum of about 9\% (see
Figures \ref{f:redblue} and \ref{f:SIKUM_Iks_change_B_16nT_f16}). A
similar measurement of the potassium current in the KCNQ1 channel
protein, expressed in an oocyte, gave a maximal increase of 16\% at
the same applied EMF and at membrane depolarizations between -10 mV
and -3 mV (see Figure \ref{f:SIKUM_KCNQ1_change_B_16nT_f16}).
Similar experiments with L-type calcium channels showed no response
to the electromagnetic field at any frequency between 0.05 and 50
Hz.

\begin{figure}
\mbox{
\begin{minipage} {\textwidth}
\begin{center}
\begin{tabular}{c}
\psfig {figure=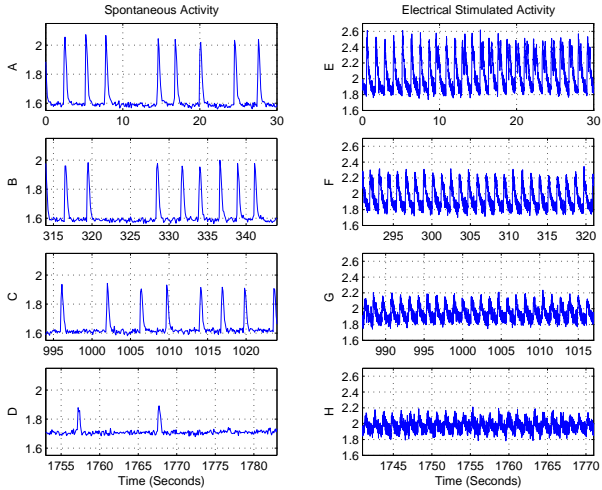,width=8cm}\\
\end{tabular}
\end{center}
\end{minipage}}
\\
\caption{\small Cardiac cells, 4 days in culture, were exposed to
magnetic fields of magnitude 160 pT and frequency 16 Hz for 30 min.
Characteristic traces of  spontaneous cytosolic calcium activity
(A,B,C,D) and of electrically stimulated (1 Hz) cytosolic calcium
activity (E,F,G,H). Times are measure in seconds from the moment of
application of the magnetic field. } \label{f:Asher1}
\end{figure}

In a related experiment \cite{Asher}, we applied electromagnetic
fields at frequencies 15 Hz, 15.5 Hz, 16 Hz, 16.5 Hz and amplitudes
of the magnetic field from below 16 pT and up to 160 nT, to neonatal
rat cardiac myocytes in cell culture. In the range 16 pT -- 16 nT,
we observed that both stimulated and spontaneous activity of the
myocytes changed at frequency 16 Hz: the height and duration of
cytosolic calcium transients began decreasing significantly about 2
minutes after the magnetic field was applied and kept decreasing for
about 30 minutes until it stabilized at about 30\% of its initial
value and its width decreased to approximately 50\%. About 10
minutes following cessation of the magnetic field the myocyte
(spontaneous) activity recovered with increased amplitude, duration,
and rate of contraction. Outside this range of frequencies and
magnetic fields no change in the transients was observed (see Figure
\ref{f:Asher1}). When the stereospecific inhibitor of KCNQ1 and
$I_{\mbox{\scriptsize Ks}}$ channels chromanol 293B was applied, the
phenomenon disappeared, which indicates that the
$I_{\mbox{\scriptsize Ks}}$ and KCNQ1 potassium channels in the
cardiac myocyte are the targets of the electromagnetic field, in
agreement with the former experiment. The effect of changing the
outward potassium current in a cardiac myocyte is to change  both
the height and duration of calcium transients, action potential,
sodium current, as indicated by the Luo-Rudy model
\cite{Zeng-Rudy1995}.

\begin{figure}
\mbox{
\begin{minipage} {\textwidth}
\begin{center}
\begin{tabular}{c}
\epsfig {figure=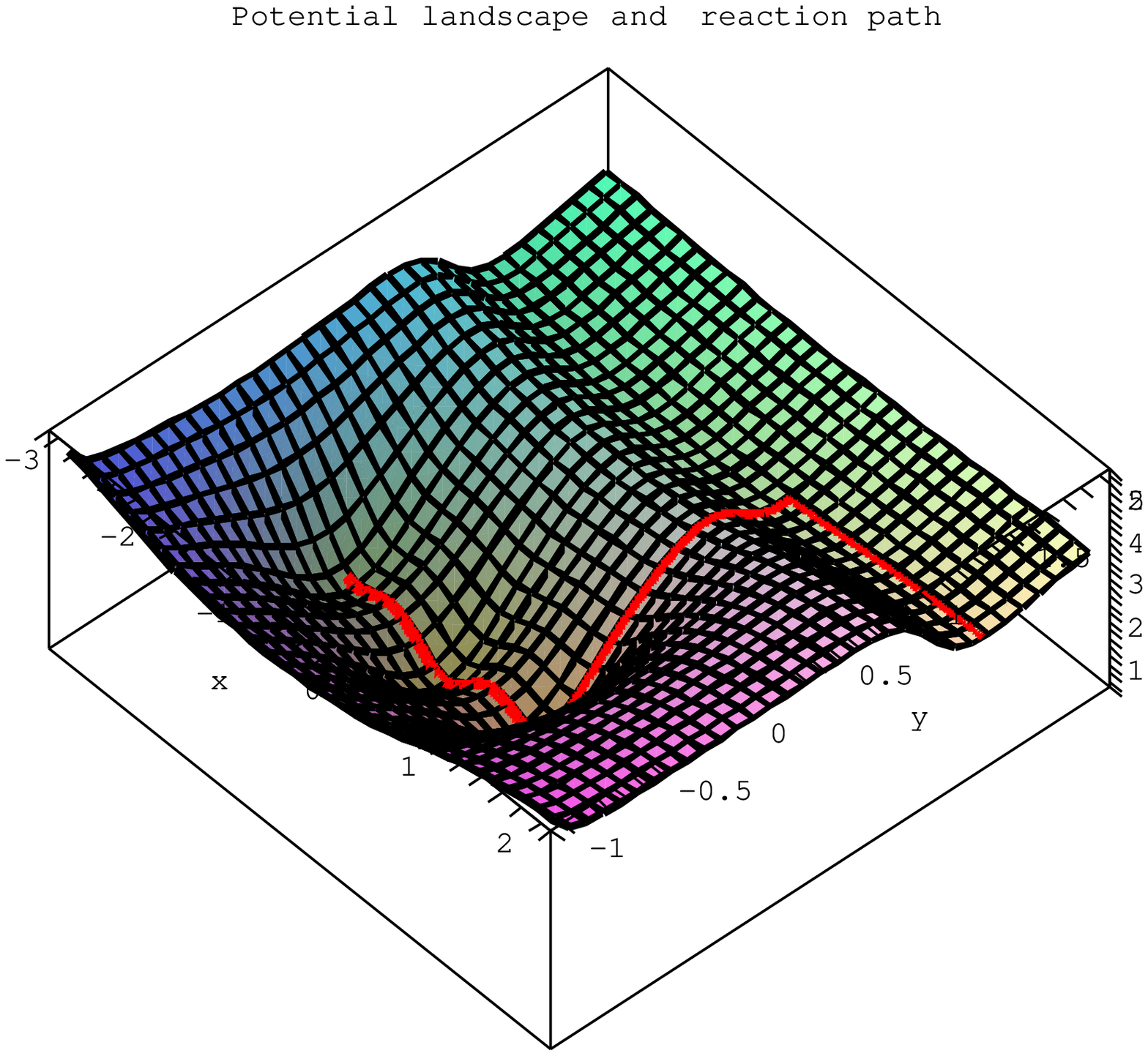,width=8cm}\\
\end{tabular}
\end{center}
\end{minipage}}
\\
\caption{\small Hypothetical energy landscape of two ions in the
selectivity filter. The reaction path is marked red. The straight
segment in the trough may represent the open state in the channel}
 \label{f:potential-path1}
\end{figure}

\begin{figure}
\mbox{
\begin{minipage} {\textwidth}
\begin{center}
\begin{tabular}{c}
\epsfig {figure=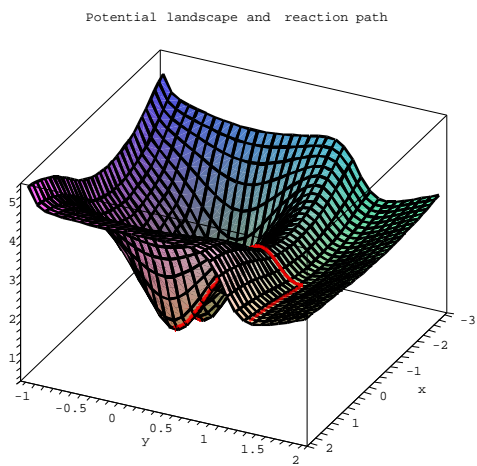,width=8cm}\\
\end{tabular}
\end{center}
\end{minipage}}
\\
\caption{\small Another view of the hypothetical energy landscape of
two ions in the selectivity filter.}
 \label{f:potential-path2}
\end{figure}

\begin{figure}
\mbox{
\begin{minipage} {\textwidth}
\begin{center}
\begin{tabular}{c}
\psfig {figure=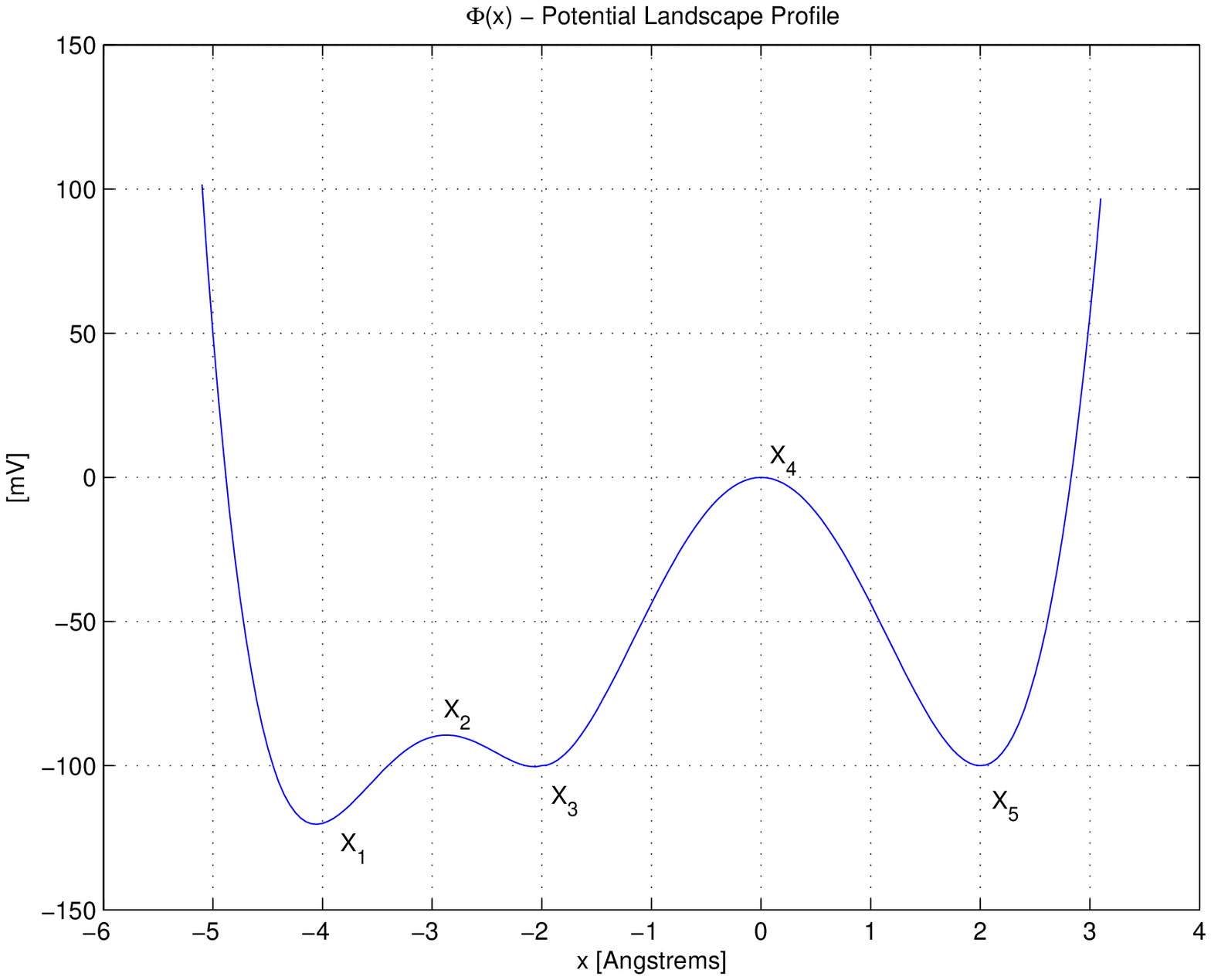,width=7cm}\\
\end{tabular}
\end{center}
\end{minipage}}
\\
\caption{\small Profile of one-dimensional electrostatic potential
landscape biased by a constant electric field}
\label{f:doublewell-landscape}
\end{figure}

\begin{figure}
\centering 
{\includegraphics[width=7cm]{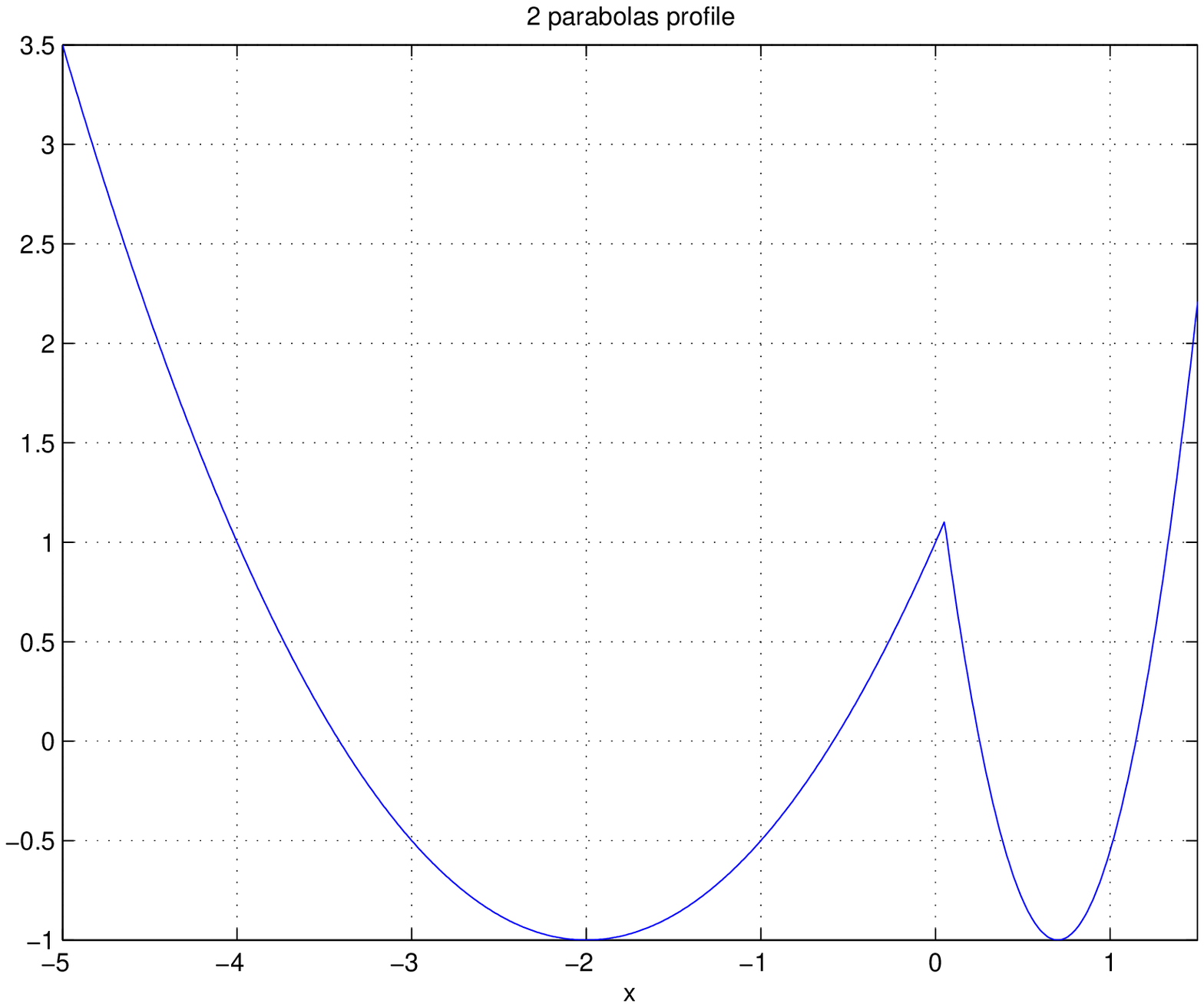}} \caption{A
simplified version (see eq.(\ref{phi0})) of the wells in Figure
\ref{f:doublewell-landscape}. The wells at $x_1$ and $x_3$ and the
barrier at $x_2$ are now at $x_1=-2, x_3=0.7$ and $x_2=0$. The
constant bias in (\ref{phixt}) is $c=0$.}\label{f:potential}
\end{figure}

The specific response at 16 Hz may indicate some form of resonance
or stochastic resonance of a gating mechanism of open voltage-gated
potassium channels (e.g., a secondary structure or mechanism) with
time-periodic induced electric field. Since the induced electric
field is too low to interact with any component of the
$I_{\mbox{\scriptsize Ks}}$ channel, we conjecture that the induced
field may interact with locally stable (metastable) configurations
of ions inside the selectivity filter  \cite{Roux}. We propose an
underlying scenario for this type of interaction based on the
collective motion of three ions in the channel, as represented in
the molecular dynamics simulation of \cite{Roux}. The configurations
of three potassium ions in the KcsA channel is represented in
\cite{Roux} in reduced reaction coordinates on a three-dimensional
free energy landscape. In our simplified model, we represent the
collective motion of the three ions in the channel as diffusion of a
higher-dimensional Brownian particle in configuration space. An
imitation hypothetical energy landscape with a reaction path
(indicated in red) is shown in Figures \ref{f:potential-path1} and
\ref{f:potential-path2}. Projection onto a reaction path reduces
this representation to Brownian motion on one-dimensional landscape
of potential barriers (see Figure \ref{f:doublewell-landscape}). The
stable states represent instantaneous crystallization of the ions
into a metastable configuration, in which no current flows through
the channel, that is, they represent closed states of the channel.
There is also a pathway in the multidimensional energy landscape
that corresponds to a steady
\newpage
\noindent current flowing in the channel, e.g., an unobstructed
trough in the energy landscape. Transitions from the latter into the
former represent gating events. In our scenario the motion between
closed states is simplified to one-dimensional Brownian motion,
e.g., in a trough obstructed  with barriers, while the interruptions
in the current correspond to exits from the unobstructed trough into
the obstructed one. Activated transitions over barriers separating
two closed states in the obstructed trough (see Figure
\ref{f:potential}) affect the probability of transition from closed
to open states. Stochastic resonance between two closed states may
change the transition rates between them, thus affecting the open
(or closed) probability of the channel (see Section \ref{s:Markov}).

We investigate the stochastic resonance (SR) in our mathematical
model of a Brownian particle in an {\em asymmetric} bistable
potential with an {\em induced} electric field. The difference
between this problem and that of the extensively studied SR with an
{\em applied} periodic electric field \cite{Gammaitoni1998},
\cite{Bulsara2003} is that according to Faraday's law (or Maxwell's
equations), the amplitude of
\noindent the induced field is proportional to the frequency of the
applied magnetic field. While the traditional manifestation of SR is
a peak in the power spectral density of the trajectory of the
resonating particle, we consider its manifestation in the
probability to be in one of the two meta-stable states. This measure
of SR is ineffective for a symmetric potential, because this
probability is $1/2$ in the symmetric case and is independent of the
applied periodic field. It is effective, however, in asymmetric
potentials, for example, when a constant bias field \noindent
depolarizes the membrane, as is the case in the above mentioned
experiments. Note that in the second experiment the depolarization
of the myocyte membrane is due to the action potential in the cell.
In contrast, asymmetry of the potential can weaken SR with an
applied field, as shown in \cite{WioBouzat1999}, \cite{Li2002}.

\begin{figure}
\centering
{\includegraphics[width=8cm]{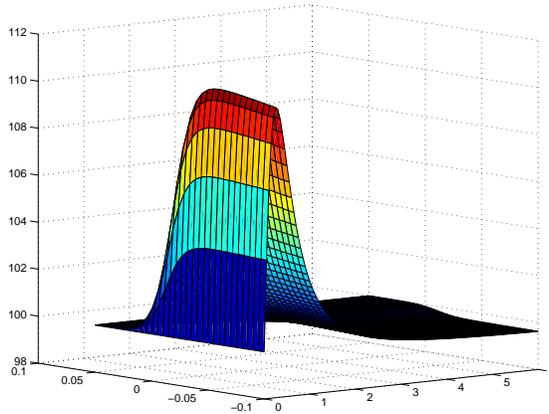}}
\caption{$P_o(\omega,c)/P_o(0,c)$ with induced force $Aw$,
$A=0.007,\ \eps = 0.029,\ x_L = -2.4,\ x_R = 1.385 $}
\label{f:Kramers_Normalized_Open_Porbability_2D_VS CW}
\end{figure}

Our main results concern SR with applied and with induced external
periodic forces. In the former case, which we view as a benchmark
for our method of analysis, we find that there is no SR as frequency
and depolarization are varied, in agreement with known results
\begin{figure}
\centering
{\includegraphics[width=8cm]{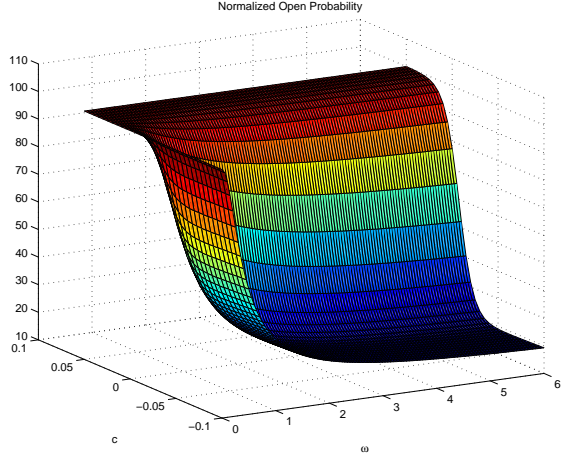}}
\caption{$P_o(\omega,c)/P_o(0,c)$ with induced force $A$, $A=0.007,
\eps = 0.029,\ x_L = -2.4,\ x_R = 1.385$  for
$0<\omega<6,-0.1<c<0.1$. Evidently, there is no SR.}
\label{f:applied1}
\end{figure}
\noindent \cite{Gammaitoni1998} (see Figure \ref{f:applied1}). In
contrast, the probability to be on one side of the barrier in the
case of an induced field peaks at a nearly fixed frequency in a
finite window of depolarizations (see Figure
\ref{f:Kramers_Normalized_Open_Porbability_2D_VS CW}). We refer to
this peak as stochastic resonance, though it may not be the usual SR
phenomenon. The folding of the surface in Figure \ref{f:applied1}
into that in Figure \ref{f:Kramers_Normalized_Open_Porbability_2D_VS
CW} seems to be due to the decrease in the amplitude of the induced
field at low frequencies. This observation is consistent with the
above mentioned experiments and seems to be new.

\begin{figure}
\mbox{
\begin{minipage} {\textwidth}
\begin{center}
\begin{tabular}{c}
\psfig {figure=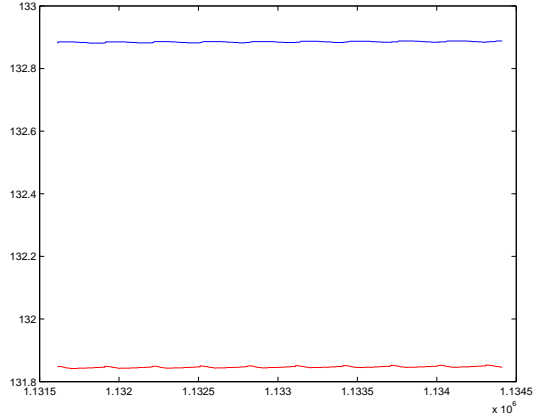,width=7cm}\\
\end{tabular}
\end{center}
\end{minipage}}
\\
\caption{\small Cytosolic potassium concentration [mM] vs time
[msec] without SR (blue) and with SR (red) in the Luo-Rudy model.
{\bf SPECIFY UNITS}} \label{f:intracelluar_potasium}
\end{figure}

\begin{figure}
\mbox{
\begin{minipage} {\textwidth}
\begin{center}
\begin{tabular}{c}
\psfig {figure=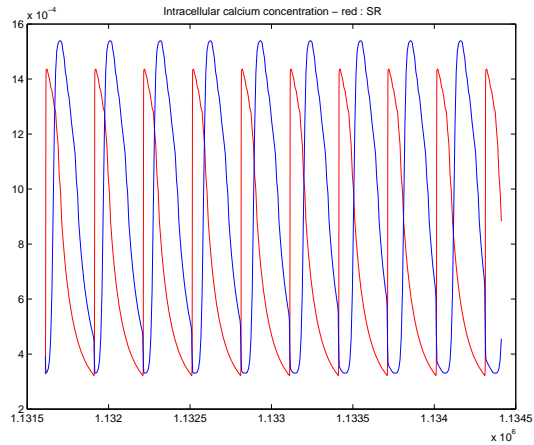,width=7cm}\\
\end{tabular}
\end{center}
\end{minipage}}
\\
\caption{\small Cytosolic calcium concentration [mM] vs time [msec]
without SR (blue) and with SR (red) in the Luo-Rudy model. }
\label{f:intracelluar_calcium}
\end{figure}

\begin{figure}
\mbox{
\begin{minipage} {\textwidth}
\begin{center}
\begin{tabular}{c}
\psfig {figure=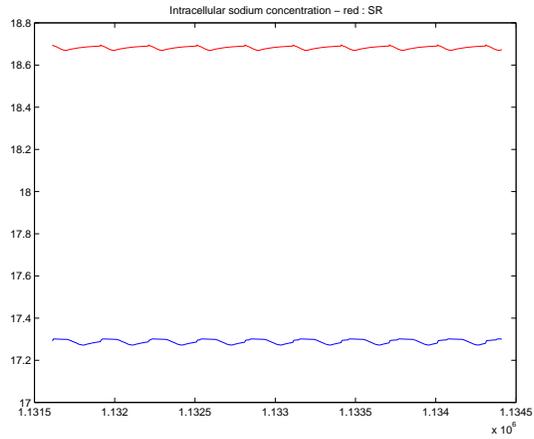,width=7cm}\\
\end{tabular}
\end{center}
\end{minipage}}
\\
\caption{\small Cytosolic sodium concentration concentration [mM] vs
time [msec] without SR (blue) and with SR (red) in the Luo-Rudy
model. } \label{f:intracelluar_sodium}
\end{figure}

\begin{figure}
\mbox{
\begin{minipage} {\textwidth}
\begin{center}
\begin{tabular}{c}
\psfig {figure=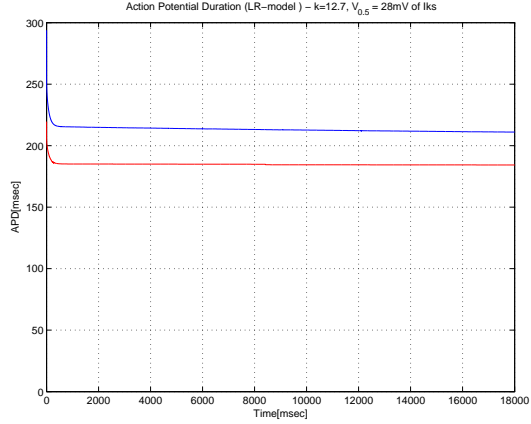,width=7cm}\\
\end{tabular}
\end{center}
\end{minipage}}
\\
\caption{\small Action potential duration [msec] vs time [msec]
without SR (Blue) and with SR (Red) in the Luo-Rudy model.}
 \label{f:Action potential duration}
\end{figure}

To connect the above SR with the cardiac myocyte experiment, we use
the Luo-Rudy model \cite{Zeng-Rudy1995} of a ventricular cardiac
myocyte of a Guinea pig. We express the manifestation of the above
SR in the Hodgkin-Huxley equations \cite{HH1952} as a change in the
conductance of the $I_{\mbox{\scriptsize Ks}}$ channel in the
specific range of depolarizations at the resonant frequency of 16
Hz. We note that the $I_{\mbox{\scriptsize Ks}}$ is one of the
delayed rectifier K$^+$ channels that are present in cardiac
myocytes \cite{Opie}, in neuron cells \cite{Koch1999},
\cite{Johnston}, and more, that is, it stays open long enough for
its (secondary) gating to partially synchronize with the induced
field. The SR-increased efflux of potassium (see Figure
\ref{f:intracelluar_potasium}) shortens the action potential, and
consequently lowers the peak of the cytosolic calcium concentration
(see Figure \ref{f:intracelluar_calcium}), at the expense of
increased sodium concentration (see \ref{f:intracelluar_sodium}).
The shortening of the action potential leads to the shortening of
the QT interval (see Figures \ref{f:Action potential myocyte},
\ref{f:Action potential duration}) \cite{Opie} and was actually
observed experimentally \cite{Mazhari et al.}, \cite{Jeong et al.}.
These predictions of the SR modified Luo-Rudy equations are also
new. In addition, we obtain from the SR modified Luo-Rudy model an
increased conductance during the plateau of the action potential in
the cardiac myocyte. This in turn shortens both the action potential
and the cytosolic calcium concentration spike durations, lowers
their amplitudes, increases cytosolic sodium, and lowers cytosolic
potassium concentrations. These theoretical predictions are
supported by experimental measurements. Specifically, these effects
were communicated in \cite{Mazhari et al.}, \cite{Jeong et al.}, as
well as in our own measurements \cite{Asher}.
\begin{figure}
\mbox{
\begin{minipage} {\textwidth}
\begin{center}
\begin{tabular}{c}
\psfig {figure=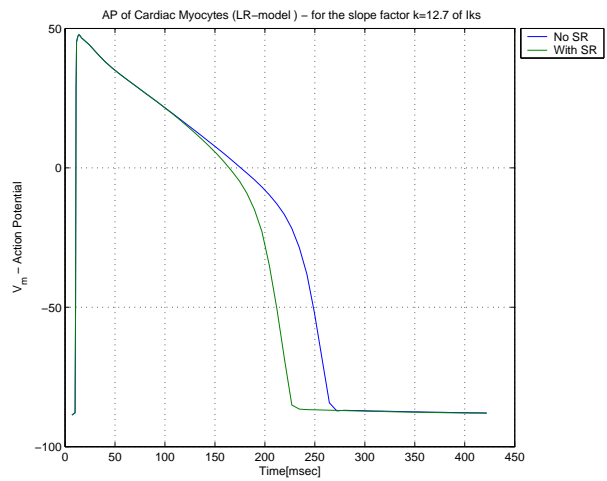,width=8cm}\\
\end{tabular}
\end{center}
\end{minipage}}
\\
\caption{\small Action potential [mV] vs time [msec] without SR
(Blue) and with SR (Green) in the Luo-Rudy model} \label{f:Action
potential myocyte}
\end{figure}

\newpage
\section{The mathematical model}\label{sec:Form}
\begin{figure}
\centering
{\includegraphics[width=8cm]{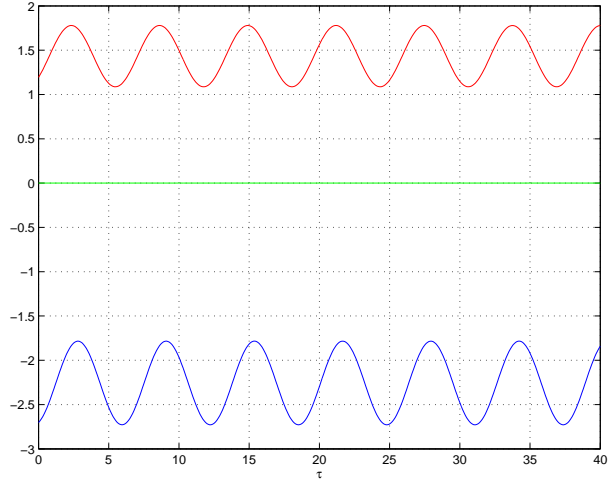}} \caption{The deterministic
trajectories  (in dimensionless units) with $\omega=1,\ A=0.5, \
c=-0.05,\ x_L = -2.4,\ x_R = 1.385 $ are attracted to the periodic
$\tilde x_L(t)$ (blue lower curve), $\tilde x_M(t)$ (green middle
curve) and $\tilde x_R(t)$ (red upper curve).} \label{f:det-traj}
\end{figure}

We consider the dimensionless overdamped dynamics
 \beq
 \dot x=-\frac{\p\phi(x,t)}{\p x}=-\phi_x(x,t)\label{dotx}
  \eeq
in the bistable time-periodic potential
 \beq
  \phi(x,t)=(c-A_{\mbox{\scriptsize Appl,Ind}})\sin\omega
  t)x+\phi_0(x),\label{phixt}
 \eeq
where $A_{\mbox{\scriptsize Appl,Ind}}$ is the amplitude of the
applied (induced) electric field and $\phi_0(x)$ is a fixed
parabolic double well potential that consists of the two parabolas
  \beq
\phi_0(x)=\left\{\begin{array}{lll}\ds \frac{(x-x_L)^2}{x_L^2}-1&\mbox{for}&x<0\\
&&\\
\ds \frac{(x-x_R)^2}{x_R^2}-1&\mbox{for}&x>0,
\end{array}\right.\label{phi0}
 \eeq
where $x_L<0<x_R$. The amplitude of the electric field induced by
the time-periodic magnetic field $B\cos\omega t$ ($B=const$) is
$A_{Ind}=A\omega$, where $A=CB$ and $C$ is the proportionality
constant in Faraday's law. The linear term $cx$ represents the
membrane depolarization. This model can be considered the limit of
the parabolic double well potential that consists of the three
parabolas
 \beq
\phi_0(x)=\left\{\begin{array}{lll}\ds\frac{(x-x_L)^2}{x_L^2}
-1+\ds\frac{1}{1+ax_L^2/2}&\mbox{for}&x<-x_{\delta_L}\\
&&\\
-\ds\frac{ax^2}{2}&\mbox{for}&-x_{\delta_L}<x<x_{\delta_R}\\
&&\\
\ds\frac{(x-x_R)^2}{x_R^2}-1+\ds\frac{1}{1+ax_R^2/2}&\mbox{for}&x>x_{\delta_R},
\end{array}\right.\label{3par}
 \eeq
where $x_L<-x_{\delta_L}<0<x_{\delta_R}<x_R$ and $a>0$. The three
parabolas connect smoothly at $-x_{\delta_L}$ and $x_{\delta_R}$,
which implies the relationships $
x_{\delta_L}=-\ds\frac{x_L}{1+ax_L^2/2}$,
$x_{\delta_R}=\ds\frac{x_R}{1+ax_R^2/2}$,
$\ds\lim_{a\to\infty}ax_{\delta_L}=-2/x_L$ and
$\ds\lim_{a\to\infty}ax_{\delta_R}= 2/x_R$. The potential
$\phi(x,t)$ (see Figure \ref{f:potential}) has two periodic
attractors, $\tilde x_L(t)$ and $\tilde x_R(t)$ (see Figure
\ref{f:det-traj})
and the separatrix\footnote{In the model (\ref{3par}) of three
parabolas the separatrix is $ \tilde x_M(t)=\frac{c}{a}-\left(
\frac{A\omega }{a^2 +\omega^2}\right)[\omega\cos\omega t+a\sin\omega
t]$.} $\tilde x_M(t)=0$. The attractors are the stable periodic
solutions of (\ref{dotx}), given by
 \beq
\tilde x_i(\tau) &=& \alpha_i- \tilde A_i \cos(\tau + \tilde\varphi_i),\quad i=L,R\label{xm_parabolas}\\
&&\nonumber\\
\tilde A_i&=&\frac{A_{Appl,Ind}x_i^2}{\sqrt{4+x_i^4\omega^2}},
\quad\tilde\varphi_i=\arctan\frac{2}{x_i^2\omega},\quad\alpha_i=\ds\left(x_{i}
- \frac{cx_i^2}{2}\right). \label{tildeA_parabola}
 \eeq

When small white noise $\sqrt{2\varepsilon}\,\dot w(t)$ is added to
the dynamics (\ref{dotx}), it becomes the stochastic equation
 \beq
 \dot x=-\phi_x(x,t)+\sqrt{2\varepsilon}\,\dot w(t).\label{sde}
  \eeq
The trajectories of (\ref{sde}) spend relatively long periods of
time near the attractors $\tilde x_L(t)$ and $\tilde x_R(t)$,
crossing $\tilde x_M(t)$ at random times. The first passage time
from $\tilde x_L(t)$ to $\tilde x_R(t)$ is defined as
 \beq
 \tau_L(t_0)=\inf\{t>0\,:\,x(t_0)={\tilde x}_L(t_0),\, x(t_0+t)={\tilde x}_R(t_0+t)\}
 \eeq
and the mean first passage time is defined as
 \beq
 \bar\tau_L=\frac{1}{T}\int_0^T\eE\tau_L(t_0)\,dt_0,
 \eeq
where $\eE$ denotes ensemble averaging over trajectories of
(\ref{sde}) and the period is
 \beqq
 T=\ds\frac{2\pi}{\omega}.
 \eeqq
The first passage time $\tau_R(t_0)$ and the mean first passage time
$\bar\tau_R$ are defined in an analogous manner. The fraction of
time the random trajectory $x(t)$ spends in the basin of attraction
of ${\tilde x}_R(t)$, that is, the fraction of time that $x(t)>
{\tilde x}_M(t)$, is the {\em right probability}
$P_R(c,\omega,A_{Appl,Ind},\varepsilon) $, given by
 \beq
P_R(c,\omega,A_{Appl,Ind},\varepsilon)
&=&\lim_{n\to\infty}\frac{1}{nT}\int_0^{nT}\int_{{\tilde
x}_M(t)}^\infty p(x,t)\,dx\,dt\nonumber\\
&&\nonumber\\
&=&\lim_{n\to\infty}\frac{1}{T}\int_0^{T}\int_{{\tilde
x}_M(t+nT)}^\infty p(x,t+nT)\,dx\,dt\nonumber\\
&&\nonumber\\
&=&\frac{1}{T}\int_0^{T}\int_{{\tilde x}_M(t)}^\infty
p_{\infty}(x,t)\,dx\,dt,\label{P_R}
 \eeq
where $p(x,t)$ is the transition probability density function (pdf)
of the random process $x(t)$, generated by the stochastic dynamics
(\ref{sde}) and $p_{\infty}(x,t)=\ds\lim_{n\to\infty}p(x,t+nT)$ is
the periodic pdf. We obtain in a similar manner
 \beq
P_L(c,\omega,A_{Appl,Ind},\varepsilon)
=\frac{1}{T}\int_0^{T}\int_{-\infty}^{{\tilde x}_M(t)}
p_{\infty}(x,t)\,dx\,dt,\label{P_L}
 \eeq
For small $\varepsilon$,
 \beq
P_L(c,\omega,A_{Appl,Ind},\varepsilon) \approx
\ds\frac{\bar\tau_{L}}{\bar\tau_{R} +
\bar\tau_{L}}.\label{Popen_MFPT}
 \eeq

\section{The Fokker-Planck equation}\label{s:FPE}
The $T$-periodic pdf $p_\infty(x,t)$ is the $T$-periodic solution of
the Fokker-Planck equation
 \beq
\frac{\p p(x,t)}{\p t}&=&\varepsilon\frac{\p^2p(x,t)}{\p x^2}+
\frac{\p[\phi_x(x,t)p(x,t)]}{\p x}
 \quad\mbox{for}\quad -\infty<x<\infty,\ 0<t<\infty.\label{FPE}
 \eeq
We construct a WKB approximation to $p_\infty(x,t)$ for small
$\eps$,
 \beq
p_\infty(x,t)\sim\frac{\exp\left\{-\ds\frac{\psi(x,t,\varepsilon)}
{\varepsilon}\right\}}{\ds\int_{-\infty}^\infty
\exp\left\{-\ds\frac{\psi(x,t,\varepsilon)}{\varepsilon}\right\}\,dx},\label{periodicWKB}
 \eeq
where $\psi(x,t,\varepsilon)$ is a $T$-periodic regular function of
$\varepsilon$. Expanding
 \beq
 \psi(x,t,\eps)=\psi(x,t,0)+\eps\psi_1(x,t)+\ldots,\label{Taylor}
  \eeq
we find from large deviations theory that $\psi(x,t,0)$ is the+
minimum of the integral
 \beq
 I(x(\cdot))(x,t)=\int_0^t\left[\dot
 x(s)+\phi_x(x(s),s)\right]^2\,ds
 \eeq
over all continuous trajectories $x(\cdot)$ such that $x(0)=x$.
Setting $\tau=\omega t$, we write the Hamilton-Jacobi (eikonal)
equation for the minimal values of $I(x,\tau)$ in the domains $x>0$
and $x<0$ as
 \beq
-\psi_{\tau}^{i}(x,\tau,0)=\frac{1}{\omega}\ds\left(\psi_x^i\right)^2(x,\tau,0)
-\frac{1}{\omega}\ds\left[c-A_{Appl,Ind}\sin\tau +
\frac{2(x-x_{i})}{x_i^2}\right]
\psi_{x}^{i}(x,\tau,0),\label{eikonalTau_parabola}
 \eeq
for $i=L,R$. The solution can be constructed in the quadratic form
\cite{GT}
 \beq \psi^i(x,\tau,0) = \ds\frac{[x -
\tilde x_i(\tau)]^2}{x_i^2} + a_i
 \eeq
and the constants $a_i$ are determined from the Freidlin-Wentzell
extremum principle \cite{FW}. According to this principle the local
minima $\psi^{L}(x,\tau,0)$ and $\psi^{R}(x,\tau,0)$ are joined into
a global minimum function $\psi(x,\tau,0)$ by the requirement that
the steady state probability current across the separatrix $\tilde
x_M$ vanishes \cite{GT},
 \beq
  J(0,\tau) =
\int_0^{2\pi} [J_L(0,\tau) + J_R(0,\tau)]\,d\tau =0,\label{currentJ}
 \eeq
where the probability flux density is
 \beq
 J_i(0,\tau)=-\eps\frac{\p p^i(0,\tau)}{\p x} +
\phi_x(0,\tau)p^i(0,\tau)= -\eps\frac{\p p^i(0,\tau)}{\p
x}.\label{current-prob1}
 \eeq
Using the WKB approximation (\ref{periodicWKB}) for $p(x,\tau)$, we
find that the minimum condition is
 \beq
  J_i(0,\tau) =\frac{-2 \tilde
x_i(\tau)}{x_i^2}\exp\left\{-\frac{1}{\eps}\left(\ds\frac{ \tilde
x_i^2(\tau)}{x_i^2}
 + a_i\right)\right\}.\label{Ji0}
 \eeq
Using  (\ref{Ji0}) in (\ref{currentJ}), we find that
 \beq
e^{(a_L - a_R)/\eps} =- \ds\frac{x_R^2\ds\int_0^{2\pi}\tilde
x_L(\tau)\exp\left\{-\frac{\tilde x_L^2(\tau)}{\eps
x_L^2}\right\}\,d\tau } {x_L^2\ds\int_0^{2\pi}\tilde
x_R(\tau)\exp\left\{-\frac{\tilde x_R^2(\tau)}{\eps
x_R^2}\right\}\,d\tau}.\label{constants_a3Cond1}
 \eeq
It should be noted that the phases $\tilde\varphi_L$ and
$\tilde\varphi_R$ may be disregarded in the integrals of equation
(\ref{constants_a3Cond1}), therefore we set them to zero. Expanding
the integrals in (\ref{constants_a3Cond1}) by the Laplace method for
small $\eps$ about the maxima of the integrands, at
 \beq
x_1 = \tilde x_R(0) = \alpha_R - \tilde A_R,\quad x_2 = \tilde
x_L(\pi) = \alpha_L - \tilde A_L,\label{x1x2}
 \eeq
we get
 \beq
e^{(a_L - a_R)/\eps} &=& - \ds\frac{x_R^2 \sqrt{\ds\frac{2\pi\eps
x_L^2}{[\tilde x_L^2]^{''}(\pi)} } \tilde
x_L(\pi)\exp\left\{-\ds\frac{\tilde x_L^2(\pi)}{\eps
x_L^2}\right\} } {x_L^2 \sqrt{\ds\frac{2\pi\eps x_R^2}{[\tilde
x_R^2]^{''}(0)} } \tilde x_R(0)\exp\left\{-\ds\frac{\tilde
x_R^2(0)}{\eps
x_R^2}\right\} }\nonumber\\
&&\nonumber\\
&=& - \ds\frac{x_R^2 \sqrt{\ds\frac{x_L^2}{-2\tilde A_Lx_2} }
x_2\exp\left\{-\ds\frac{ x_2^2}{\eps x_L^2}\right\} } {x_L^2
\sqrt{\ds\frac{x_R^2}{2\tilde A_Rx_1} } x_1 \exp\left\{-\ds\frac{
x_1^2}{\eps x_R^2}\right\} },\label{constants_a3Cond2}
 \eeq
 where, according to (\ref{xm_parabolas}) and $\tilde\varphi_i =
 0$

\subsection{The left probability
$P_{L}(c,\omega,A_{Appl,Ind},\varepsilon)$}\label{s:SR_PL}

To calculate the probability
$P_{L}(c,\omega,A_{Appl,Ind},\varepsilon)$, we use the WKB
approximation (\ref{periodicWKB}) in (\ref{P_L}) and evaluate the
integrals by the Laplace method, as in (\ref{Ji0}), to get
 \beq
P_{L}(c,\omega,A_{Appl,Ind},\varepsilon)=\ds\frac{1}{1+\ds\frac{x_R}{-x_L}e^{(a_L
- a_R)/\eps}} .\label{Popen}
 \eeq
Using the result from (\ref{constants_a3Cond2}) in (\ref{Popen}), we
find that
 \beq
P_{L}(c,\omega,A_{Appl,Ind},\varepsilon)
=\ds\frac{1}{1+\ds\frac{x_R^3}{|x_L|^3}\sqrt{\ds\frac{\sqrt{4+x_L^4\omega^2}}{\sqrt{4+x_R^4\omega^2}}}
\sqrt{\ds\frac{-x_2}{x_1}}\exp\left\{
\ds\frac{ x_1^2}{\eps x_R^2}-\ds\frac{ x_2^2}{\eps
x_L^2}\right\}}.\label{Popen1}
 \eeq
We normalize the frequency-dependent left probability
$P_{L}(c,\omega,A_{Appl,Ind},\varepsilon)$ by the left probability
of the unforced dynamics
$P_{L}^{0}(c,\omega,A_{Appl,Ind}=0,\varepsilon)$. To calculate
$P_{L}^{0}(c,\omega,A_{Appl,Ind}=0,\varepsilon)$, we set $A=0$ in
(\ref{Popen}) and obtain
 \beq
P_{L}^{0}(c,\omega,A_{Appl,Ind}=0,\varepsilon)&=&
\left[1+\frac{x_R^2}{x_L^2} \ds\left| \ds\frac{2-cx_L}{2-cx_R}
\right|\,
\exp\left\{\frac{c(x_R-x_L)}{\eps}\left(\frac{c(x_R+x_L)}{4}-1\right)\right\}
\right]^{-1}.\nonumber\\
&&\nonumber\\\label{Popen_A0}
 \eeq
 Note that (\ref{Popen_A0}) is not other than (\ref{Popen_MFPT}),
 where
 \beq \bar\tau_i =
\ds\sqrt\frac{2\pi\varepsilon}{\phi_{xx}^{i}(\tilde{x}_i)|\phi_{x}^{i}(\tilde{x}_M)\,|^2}\ds\exp\left(
\ds\frac{\phi^{i}(\tilde{x}_M) -
\phi^{i}(\tilde{x}_i)}{\varepsilon}\right).\label{MFPT_sharpBoundary}
 \eeq
The MFPT $\bar\tau_i$ in (\ref{MFPT_sharpBoundary}) is the Kramers
escape rate of a Brownian particle over a high sharp barrier
$\tilde{x}_{M}$ \cite{book}.
\section{Coarse-grained Markov model of secondary gating}\label{s:Markov}

We consider the movement of a Brownian particle over two unequal
barriers of heights $\Delta\phi_{21} = \phi(x_{2}) - \phi(x_{1})$
and $\Delta\phi_{43} = \phi(x_{4}) - \phi(x_{3})$, respectively,
such that $\Delta\phi_{21} \ll \Delta\phi_{43}$ (see Figure
\ref{f:doublewell-landscape}). Both $x_{1}$ and $x_{3}$ are closed
states of the channel whereas $x_{5}$ represents the open state (see
Figure \ref{f:potential-path1}). Our goal is to elucidate the
influence of SR between the periodic force and the activation over
the local small barrier $\Delta\phi_{21}$ at $x_{2}$, within the
closed state, on the open probability of the channel. Specifically,
when SR increases the time spent in the well at $x_3$ relative to
that at $x_1$, the attempt frequency to cross the barrier at $x_4$
into the open state $x_5$ increases, thus increasing the open
probability of the channel. More specifically, we evaluate the
influence of SR on the mean closed time, that is, on the mean time
spent in the wells at $x_1$ and $x_3$ prior to passage into $x_5$
(which we denote $\bar{\tau}_{1,3}^{c}$). For that purpose, we can
assume that $x_{5}$ is an absorbing boundary.

First, we note that steady state considerations can be applied in
describing SR in the wells at $x_1$ and $x_3$. Indeed, we assume
that
 \beq
\phi(x_4)-\phi(x_5),\phi(x_4)-\phi(x_3)>\phi(x_2)-\phi(x_1)>
\phi(x_2)-\phi(x_3)\gg\varepsilon,\label{barriers}
 \eeq
which means that a transition over the barrier at $x_4$ between the
open and closed states occurs at a much lower rate than those over
the barrier at $x_2$, between the two closed substates. In
particular, the first inequality in (\ref{barriers}) means that
there will be many transitions over the barrier at $x_2$ before a
transition occurs from $x_3$ to $x_5$ over the barrier at $x_4$.
Thus we confine our attention to transitions over the former and
consider $x_5$ to be an absorbing state, as mentioned above. The
assumption of high barriers (the last inequality in (\ref{barriers})
means that a quasi steady state is reached in each of the wells
before a transition over $x_4$ occurs. Therefore the pdf of the
quasi-steady state in each well can be represented by the principal
eigenfunction and eigenvalue in that well, with absorbing boundary
conditions.

We coarse-grain the trajectory of the diffusion process $x(t)$ into
that of a continuous-time three state Markov jump process
$\tilde{x}(t)$, that jumps between $x_{1}$ and $x_{3}$ and is
absorbed in $x_{5}$,
 \beq x_{1} \rightleftarrows x_{3} \rightarrow
x_{5}. \label{MarkovChain}
 \eeq
 The three state continuous-time Markov chain $\tilde{x}(t)$ is not stationary
due to the passage to the open state $x_5$. There are two time
scales of passages: a short scale corresponding to the transitions
between $x_{1}$ and $x_{3}$, and a long one for the transitions
between $x_{3}$ and $x_{5}$. Due to the long time scale, the
dynamics between the closed states ($x_{1}$ and $x_{3}$)is
quasi-stationary. We assume it as a stationary dynamics  in our
analysis.

 The jump of the Markov process from $x_1$ to $x_3$ occurs when
$x(t)$ reaches $x_3$ for the first time after it was at $x_1$, and
so on. The Chapman-Kolmogorov equation for the transition
probability matrix $\mb{P}_{t}$ of the Markov process is
\cite{Ross83}
\begin{equation}
\dot{\mb{P}_{t}} = \mb{P}_{t}\mb{R}, \label{ChapmanKolmogorov}
\end{equation}
where
 \beq
 P_{t}(i,j) = \Pr\{\tilde{x}(t)= j\,|\,\tilde{x}_{0}=i\},
\quad \mbox{for}\quad 1\leq i, j \leq 3, \label{PijMatrix}
 \eeq
and the elements of the instantaneous jump rate matrix $\mb{R}$ are
 \beq
 \mb{R} = \left[
\begin{array}{ccc}
-r_{13} & r_{13} & 0\\
& & \\
r_{31} & -(r_{31}+ r_{35})  & r_{35}\\
& & \\
0 & 0 & 0
\end{array}
\right].
 \eeq
The stationary distribution $\mb{\pi}$ of the process is
\[
\mb{\pi} = \left[
\begin{array}{lll}
0, & 0, & 1
\end{array}
\right],
\]
because $x_5$ is an absorbing state.

Next, we calculate the time-dependent probability distribution
 \beq
 \mb{p}_{t} = \left[
\begin{array}{lll}
\Pr\{\tilde{x}(t) = x_{1}\}, & \Pr\{\tilde{x}(t) = x_{3}\}, &
\Pr\{\tilde{x}(t) = x_{3}\}
\end{array}
\right]^{T}.
 \eeq
According to (\ref{ChapmanKolmogorov}) and (\ref{PijMatrix}),
$\mb{p}_{t}$ the sum of elements in each column of the matrix
$\mb{P}_t$ 
$$Pr\{\tilde{x}(t) = x_{j}\} = \sum_{i=1}^{3}\Pr\{\tilde{x}(t)=
j\,|\,\tilde{x}_{0}=i\} \quad \mbox{for} \quad 1\leq j \leq 3$$ and
therefor satisfies the Chapman-Kolmogorov equation
 \beq
 \dot{\mb{p}}_{t} =
\mb{R}^T\mb{p}_{t}, \label{ChapmanKolmogorovVectorial}
 \eeq
given by
 \beq
\mb{p}_{t}=
e^{\mb{R}^{T}t}\mb{p}_{0},\label{ChapmanKolmogorovVectorialSolution}
 \eeq
where we assume that $\mb{p}_{0}$ is the stationary distribution
of the chain $ x_{1} \rightleftarrows x_{3}$, namely,
 \beq
 \mb{p}_{0} =
\left[
\begin{array}{lll}
\ds{\frac{r_{31}}{r_{13}+r_{31}}}, &
\ds{\frac{r_{13}}{r_{13}+r_{31}}}, & 0\\
\end{array}
\right]^{T}.
 \eeq
We further express the vector $\mb{p}_{t}$ as a linear combination
of the eigenvectors $\{\mb{v}_{1},\mb{v}_{2},\mb{v}_{3}\}$ of the
matrix $\mb{R}^{T}$, corresponding to the eigenvalues
$\lambda_{1},\lambda_{2},\lambda_{3}$,
 \beq
 \mb{p}_{t} =
\alpha_{1} e^{\lambda_{1}t}\mb{v}_{1} + \alpha_{2}
e^{\lambda_{2}t}\mb{v}_{2} + \alpha_{3}
e^{\lambda_{3}t}\mb{v}_{3}\label{PeigenvectorExpand},
 \eeq
where
\begin{eqnarray}
\lambda_{1} &=&  \ds{\frac{-r_{13}r_{35}}{S}}\label{Lambda1}  \\
&&\nonumber \\
\lambda_{2}&=& -S + \ds{\frac{r_{13}r_{35}}{S}}\label{Lambda2}\\
&&\nonumber \\
\lambda_{3} &=& 0.
\end{eqnarray}
and $S = r_{13}+r_{31}+r_{35}$. Using the fact that
$\ds\lim_{t\rightarrow\infty} \mb{p}_{t} = \mb{\pi}$, we get that
$\alpha_{3} = 1$.

 The MFPTs $\bar\tau_{3}$ and $\bar\tau_{1}$ are related to the
exit rates $r_{31}$ and $r_{13}$ over a non-sharp boundary
\cite{MST} according to
 \beq
\bar\tau_{3} = \ds\frac{1}{2r_{31}}\nonumber\\
\bar\tau_{1} = \ds\frac{1}{2r_{13}}.\label{r13r31}
 \eeq

 In order to find mean closed time $\bar{\tau}_{1,3}^{c}$ prior to the
first arrival to $x_5$, it is enough to consider the matrix
 \beq
\tilde{\mb{R}}^{T} = \left[
\begin{array}{lc}
-r_{13} & r_{31} \\
r_{13} & -(r_{13}+r_{35})
\end{array}
\right],\label{TildeR}
 \eeq
because this time is determined by the first two elements of the
vector $\mb{p}_{t}$, \beq \mb{p}_{t}^{1,3} = \left[
\begin{array}{ll}
\Pr\{\tilde{x}(t) = x_{1}\}, & \Pr\{\tilde{x}(t) = x_{3}\}
\end{array}
\right]^{T}. \label{pt13}
 \eeq
An eigenvector expansion $\mb{p}_{t}^{1,3}$, similar to that in
(\ref{PeigenvectorExpand}), is
 \beq
 \mb{p}_{t}^{1,3} =
\alpha_{1} e^{\lambda_{1}t}\tilde{\mb{v}_{1}} + \alpha_{2}
e^{\lambda_{2}t}\tilde{\mb{v}_{2}} \label{PeigenvectorExpand12},
 \eeq
where $\lambda_{1}$ and $\lambda_{2}$ are given in (\ref{Lambda1})
and (\ref{Lambda2}), respectively, and $\tilde{\mb{v}_{1}}$ and
$\tilde{\mb{v}_{2}}$ are the eigenvectors of the matrix
$\tilde{\mb{R}}^{T}$
\begin{eqnarray*}
\tilde{\mb{v}_{1}} &=& \ds{\left[ r_{31}S,
r_{13}S-r_{13}r_{35}\right]^{T}} \\
&&\\
 \tilde{\mb{v}_{2}}&=& \ds{\left[ r_{31}S, r_{13}S - S^{2}+
r_{13}r_{35}\right]^{T}}.
\end{eqnarray*}
Using the initial condition
 \beqq
 \mb{p}_{0}^{1,3} = \left[
\begin{array}{ll}
\ds{\frac{r_{31}}{r_{13}+r_{31}}}, &
\ds{\frac{r_{13}}{r_{13}+r_{31}}} \\
\end{array}
\right]^{T},
 \eeqq
we obtain
\begin{eqnarray*}
\alpha_{1} =
\ds{\frac{S^{2}+r_{13}r_{35}}{S^{3}(r_{13}+r_{35})}}\approx
\ds{\frac{1}{(r_{13}+r_{31})^{2}}},\quad
 \alpha_{2} = \ds{\frac{r_{13}r_{35}}{-S^{3}(r_{13}+r_{31})}}.
\end{eqnarray*}
Setting $P(t) = \mb{p}_{t}^{1,3}(1,1) + \mb{p}_{t}^{1,3}(2,1)$ and
substituting the values of $\alpha_{1}$, $\alpha_{2}$,
$\tilde{\mb{v}_{1}}$ and $\tilde{\mb{v}_{2}}$ into
(\ref{PeigenvectorExpand12}), we obtain
 \beq
P(t) = e^{\lambda_{1}t} -
e^{\lambda_{2}t}\ds{\frac{(r_{13}r_{35})^2}{(r_{13}+r_{31})^{4}}}.
\label{Pt}
 \eeq
Hence
 \beq
 \bar{\tau}_{1,3}^{c} = E[\tau_{1,3}^{c}] = \int_{0}^{\infty}
 P(t)\,dt \sim \ds{\frac{1}{|\lambda_{1}|}}=
 \ds{\frac{S}{r_{13}r_{35}}} \simeq \ds{\frac{1}{r_{35}}}\ds{\left( 1 +
 \ds{\frac{r_{31}}{r_{13}}}\right)}.
 \label{tau_13ca}
  \eeq
 Using (\ref{r13r31}) in (\ref{tau_13ca}) and setting $P_3^R =
\ds\frac{\bar\tau_3}{\bar\tau_3 +\bar\tau_1 }$, we obtain that
 \beq
 \bar{\tau}_{1,3}^{c} = \ds{\frac{1}{r_{35}}}\ds{\left( 1 +
 \ds{\frac{\bar\tau_{1}}{\bar\tau_{3}}}\right)} =
 \ds{\frac{1}{r_{35}}}\frac{1}{P_3^R},
 \label{tau_13cb}
 \eeq
 We further
coarse-grain the trajectories of the process $\tilde x(t)$ into
that of a telegraph process with two states, closed state
(corresponding to $x_1$ and $x_3$) and open state (corresponding
to $x_5$) $$c\rightarrow o.$$ Denoting $\bar\tau_o$ as the mean
first passage time from $x_5$,  using equation (\ref{tau_13cb}) we
get
  \beq
P_{open} = \ds\frac{\bar\tau_o}{\bar\tau_o + \bar{\tau}_{1,3}^{c}}=
\ds\frac{\bar\tau_o}{\bar\tau_o +
\ds{\frac{1}{r_{35}}}\frac{1}{P_3^R}}.
   \label{Popen1}
  \eeq
 Applying the theory proposed in \ref{s:SR_PL}, to the dynamics
between the closed states $x_1$ and $x_3$, the SR effect increases
$P_3^R$, by using a negative depolarization, $c$ . According to
(\ref{Popen1}) an increase of $P_3^R$ causes to an increase of
$P_{open}$.

\subsection{High barrier approximation to $r_{35}$, $r_{31}$, $r_{13}$}

We consider the autonomous stochastic differential equation
\begin{eqnarray}
dx &=&-\phi'(x) \,dt+\sqrt{2\varepsilon}\,dw  \label{sp1d} \\
&&\nonumber\\
 x\left( 0\right) &=&x.  \nonumber
\end{eqnarray}
The transition rates between the wells are the probability fluxes in
the direction of the transition at the top of the barrier. Thus,
denoting by $\Phi_i(x)$ and $\lambda_i$ the principal eigenfunction
and eigenvalue in well $i\ (i=1,3)$, we have
 \beq
 r_{13}=-\varepsilon{\Phi'_1}(x_2),\quad
 r_{31}=\varepsilon{\Phi'_3}(x_2),\quad
 r_{35}=-\varepsilon{\Phi'_3}(x_4).\label{rate-flux}
 \eeq

To calculate the fluxes, we have to construct the eigenfunctions
$\Phi_i(x)$, which are the solutions of
 \beq
 \varepsilon\Phi_1''(x)+\left[\phi'(x)\Phi_1(x)\right]'&=&-\lambda_1\Phi_1(x)
 \quad\mbox{for}\quad -\infty<x<x_2\label{Phi1}\\
 &&\nonumber\\
 \Phi_1(x_2)&=&0,\quad\Phi_1(x)\to0\quad\mbox{for}\quad x\to-\infty\label{BCPhi1}\\
 &&\nonumber\\
\varepsilon\Phi_3''(x)+\left[\phi'(x)\Phi_3(x)\right]'&=&-\lambda_3\Phi_3(x)
 \quad\mbox{for}\quad -x_2<x<x_4\label{Phi3}\\
 &&\nonumber\\
 \Phi_3(x_2)&=&0,\quad\Phi_3(x_4)=0.\label{BCPhi3}
 \eeq
The asymptotic structure of the eigenfunctions is given in
\cite{book} as
 \beq
 \Phi_1(x)&\sim&-{\cal N}_1^{-1}
e^{-\phi(x)/\varepsilon}\sqrt{\frac2\pi}\int_0^{\omega_2(x-x_2)/\sqrt{\varepsilon}}
e^{-z^2/2}dz\quad\mbox{for}\quad x<x_2\label{Phi1-asympt}\\
 &&\nonumber\\
 \Phi_3(x)&\sim&{\cal N}_3^{-1}
 e^{-\phi(x)/\varepsilon}\sqrt{\frac2\pi}
 \left[\int_{\omega_4(x-x_3)/\sqrt{\varepsilon}}^{\omega_2(x-x_2)/\sqrt{\varepsilon}}
e^{-z^2/2}dz- 1\right]\quad\mbox{for}\quad x_2<x<x_4,
 \label{Phi3-asympt}
 \eeq
where $\omega_i=\sqrt{|\phi''(x_i)|},\ (i=1,2,3,4)$ and
 \beq
{\cal N}_1=
\int_{-\infty}^{x_2}\Phi_1(x)\,dx\sim\frac{\sqrt{2\pi}}{\omega_1}
 e^{-\phi(x_1)/\varepsilon},\quad
 {\cal N}_3=\int_{x_2}^{x_4}\Phi_3(x)\,dx\sim\frac{\sqrt{2\pi}}{\omega_3}
 e^{-\phi(x_3)/\varepsilon}.\label{N13}
 \eeq

According to (\ref{rate-flux}) and (\ref{Phi1-asympt})-(\ref{N13}),
 \beq
 r_{13}\sim\frac{\omega_1\omega_2}{\pi}e^{-[\phi(x_2)-\phi(x_1)]/\varepsilon},
 \quad r_{31}\sim\frac{\omega_3\omega_2}{\pi}e^{-[\phi(x_2)-\phi(x_3)]/\varepsilon}
  \quad
  r_{35}\sim\frac{\omega_3\omega_4}{\pi}e^{-[\phi(x_4)-\phi(x_3)]/\varepsilon},
  \label{rates135}
  \eeq
which are Kramers' rates for the corresponding barriers \cite{book}.

\section{Effect of SR in the Luo-Rudy model of cardiac myocytes}
The Luo-Rudy model \cite{Zeng-Rudy1995} describes ionic
concentrations and cardiac ventricular action potential by a system
of $21$ ordinary differential equations. It reflects the guinea-pig
electrophysiology by detailed Hodgkin-Huxley models of ionic
currents. The most significant currents are the slow
$I_{\mbox{\scriptsize Ks}}$ and rapid $I_{\mbox{\scriptsize Kr}}$
delayed rectifier potassium currents, a time-independent potassium
current, a plateau potassium current (ultra-rapid
$I_{\mbox{\scriptsize Kur}}$), a transient outward current, fast and
background sodium currents, L- and T-type calcium currents, a
background calcium current, calcium pumps, sodium-potassium pumps,
and sodium-calcium exchangers. In addition, the model describes
$Ca^{2+}$ handling processes, that is, calcium dynamic release from
the sarcoplasmic-reticulum and from the calcium buffers troponin,
calmodulin, and calsequestrin.

The stochastic resonance described above changes the open
probability of the $I_{\mbox{\scriptsize Ks}}$ channel, and
therefore it affects its conductance. To incorporate this effect
into the Luo-Rudy model, we modify the Hodgkin-Huxley equation for
the $I_{\mbox{\scriptsize Ks}}$ current-voltage relation by shifting
the stationary open probability of the channel in the Luo-Rudy model
\cite{Zeng-Rudy1995},
 \beq
 P_O(V)= \ds\frac1{1+\ds\exp\left\{-\ds\frac{V-1.5}{16.7}\right\}},
 \eeq
to
 \beq
 \tilde
 P_O(V)=\ds\frac1{1+\ds\exp\left\{-\ds\frac{V+4.12}{16.7}\right\}},
 \label{POSR}
 \eeq
which imitates the experimentally observed shift (see figure
\ref{f:redblue}).
\begin{figure}
\centering
{\includegraphics[width=8cm]{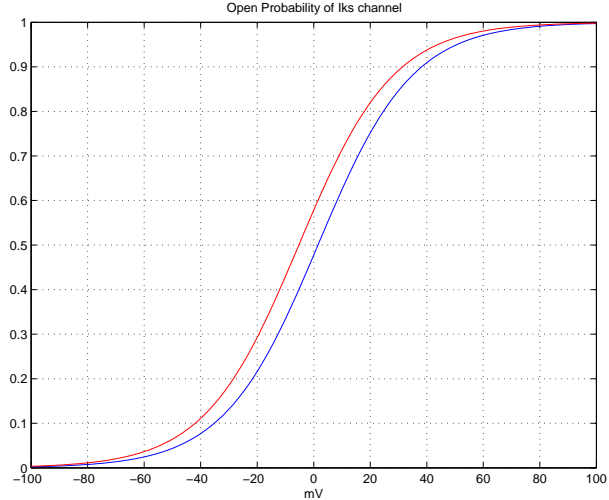}} \caption{\small Resonant
increase (red) of $20\%$  in the open probability of the
$I_{\mbox{\scriptsize Ks}}$ channel, and normal regime (blue)}
\label{f:BUMP_LRD}
\end{figure}
This changes the channel conductance $\bar G_{\mbox{\scriptsize
Ks}}P_O^2(V)$ in the Luo-Rudy model ($\bar G_{\mbox{\scriptsize
Ks}}$ is the open channel conductance) to $\bar G_{\mbox{\scriptsize
Ks}}\tilde P_O^2(V)$, which changes, in turn, the membrane potassium
current $\langle I_{\mbox{\scriptsize Ks}}\rangle$, averaged over
many channels, to \cite{Zeng-Rudy1995}
 \beq
\langle I_{\mbox{\scriptsize Ks}}\rangle\to
\bar{G}_{\mbox{\scriptsize Ks}}\tilde
P_O^2(V)(V-E_{\mbox{\scriptsize Ks}})\quad\mbox{as}\quad
t\to\infty\label{IKs}
 \eeq
($E_{Ks}$ is the reversal potential of the channel). The effect of
this modification of the Luo-Rudy model is shown in Figure
\ref{f:intracelluar_potasium}. The duration of the action potential
is reduced and accordingly, the peak of the cytosolic calcium
concentration is lowered (see Figures
 \ref{f:intracelluar_calcium}), as in the
experiment described in the Introduction. On the other hand, sodium
concentration is increased (see Figure \ref{f:intracelluar_sodium}).
The shortening of the action potential duration in the ventricular
cardiac myocytes affects the QT interval in the electrocardiogram,
which consists of a sum of several different action potentials
created in the myocardium \cite{Opie} (see Figures \ref{f:Action
potential myocyte}, \ref{f:Action potential duration}). These
theoretical predictions are supported by experimental measurements.
Specifically, these effects in vivo were communicated in
\cite{Mazhari et al.}, \cite{Jeong et al.}, as well as in our own in
vitro measurements \cite{Asher}.

\section{Conclusion and Discussion}
This paper tries to explain the results of the experiment of
exposing human potassium $I_{\mbox{\scriptsize Ks}}$ channels and
cardiac myocytes, which contain these channels, to weak and slow
electromagnetic fields. We offer a scenario of a new kind of
stochastic resonance between the induced periodic field and the
thermally activated transitions between locally stable
configurations of the mobile ions in the selectivity filter.

More specifically, since the induced electric field is too weak to
interact with any component of the $I_{\mbox{\scriptsize Ks}}$
channel protein, our model cannot describe the primary gating
mechanism of a voltage gated channel. We therefore resort to a
mathematical model, which postulates interaction of the induced
field with configurations of the mobile ions inside the selectivity
filter. These configurations may be much more susceptible to the
weak induced field than any components of the surrounding protein,
because the potential barriers separating the metastable
configurations of the mobile ions can be of any height.

According to our scenario, the observed resonance is due to the
dependence of the induced electric field amplitude on frequency, in
contrast to an applied external electric field with fixed frequency,
which is known not to exhibit stochastic resonance with changing
frequency and depolarization. In our theory the observed SR between
two closed (or inactivated) states affects the open probability of
the channel.

Our model describes the dynamics of a Brownian particle in an {\em
asymmetric} bistable potential forced by a periodic {\em induced}
electric field. The analysis of this model is based on the
construction of an asymptotic solution to the time-periodic
Fokker-Planck equation in the WKB form. We evaluate the dependence
of the steady state probability to be on one side of the potential
barrier on the frequency, amplitude, depolarization, and noise
intensity.

Our main results are shown in Figure
\ref{f:Kramers_Normalized_Open_Porbability_2D_VS CW}, which
indicates that there is a peak in the open probability in a
relatively narrow range of depolarizations and frequencies. We refer
to this peak as stochastic resonance, though it is not be the usual
SR phenomenon. This observation is consistent with the results of
the $I_{\mbox{\scriptsize Ks}}$ channel experiment mentioned in the
Introduction.

Another result is the incorporation of the SR result into the
Luo-Rudy model of cardiac myocytes. We found that the increased
conductance of the $I_{\mbox{\scriptsize Ks}}$ channel reduces the
duration of the action potential, the peak height of the cytosolic
calcium concentration, in good agreement with the experimental
results. The shortening of the action potential duration in the
ventricular cardiac myocytes affects the QT interval in the
electrocardiogram.

\noindent {\bf Acknowledgment:} We wish to thank S. Laniado, T.
Kamil and M. Scheinowitz for introducing us to the {\em in vivo}
resonance experiments, T. Zinman, A. Shainberg and S. Barzilai for
the {\em in vitro} cardiac myocytes experiments, G. Gibor and B.
Attali for the oocyte experiments, and N. Dascal and A. Moran for
experiments on L-type channels. We thank Y. Rudy, F. Bezanilla, and
G. Deutscher for useful discussions.

\end{document}